\documentclass[journal]{IEEEtran}
\IEEEoverridecommandlockouts

\usepackage{cite}
\usepackage{amsmath,amssymb,amsfonts}
\usepackage{algorithmic}
\usepackage{graphicx}
\usepackage{textcomp}
\usepackage[dvipsnames]{xcolor}
\def\BibTeX{{\rm B\kern-.05em{\sc i\kern-.025em b}\kern-.08em
    T\kern-.1667em\lower.7ex\hbox{E}\kern-.125emX}}

\usepackage{algorithm}
\usepackage{array}
\usepackage{graphicx}
\usepackage{url}
\usepackage{verbatim}
\usepackage{cite}
\usepackage{stfloats}
\usepackage{caption}
\usepackage{subcaption}
\usepackage{multirow}
\usepackage{multicol}
\usepackage{comment}
\usepackage{upgreek}

\usepackage{rotating}
\usepackage{color}
\usepackage{tabularray}
\usepackage{array}
\usepackage{caption}
\usepackage{ragged2e}
\usepackage{colortbl}
\usepackage{hhline}

\usepackage[compact]{titlesec}

\titlespacing{\section}{0pt}{2ex}{1ex}
\titlespacing{\subsection}{0pt}{1ex}{0.5ex}
\titlespacing{\subsubsection}{0pt}{0.5ex}{0ex}

\begin{document}
\title{Unsupervised SFQ-Based Spiking Neural Network}
\author{\IEEEauthorblockN{Mustafa Altay Karamuftuoglu, Beyza Zeynep Ucpinar, Sasan Razmkhah,\\
Mehdi Kamal and Massoud Pedram \vspace{-2em}}\\
\thanks{This work has been funded by the National Science Foundation under the  Expedition DISCoVER project with grant number 2124453.
(Corresponding author: M. A. Karamuftuoglu)
M.A.K., B.Z.U., S.R., M.K., and M.P. are with Ming Hsieh Department of Electrical and Computer Engineering, University of Southern California, Los Angeles, USA. (karamuft@usc.edu, ucpinar@usc.edu, razmkhah@usc.edu, mehdi.kamal@usc.edu, pedram@usc.edu)
}
}
\maketitle

\noindent
\begin{abstract}
Single Flux Quantum (SFQ) technology represents a groundbreaking advancement in computational efficiency and ultra-high-speed neuromorphic processing. The key features of SFQ technology, particularly data representation, transmission, and processing through SFQ pulses, closely mirror fundamental aspects of biological neural structures. Consequently, SFQ-based circuits emerge as an ideal candidate for realizing Spiking Neural Networks (SNNs). 
%Asynchronous SFQ circuits exhibit the remarkable ability to emulate neuronal behavior, further enhancing their suitability for SNN applications. Additionally, the bio-inspired nature of SNNs offers valuable insights into the temporal dynamics of synaptic and neuronal behaviors.
This study presents a proof-of-concept demonstration of an SFQ-based SNN architecture, showcasing its capacity for ultra-fast switching at remarkably low energy consumption per output activity. Notably, our work introduces innovative approaches: (i) We introduce a novel spike-timing-dependent plasticity mechanism to update synapses and to trace spike-activity by incorporating a leaky non-destructive readout circuit.
% Spike-Timing-Dependent Plasticity Engine: 
(ii) We propose a novel method to dynamically regulate the threshold behavior of leaky integrate and fire superconductor neurons, enhancing the adaptability of our SNN architecture.
% Dynamic Threshold Behavior: 
(iii) Our research incorporates a novel winner-take-all mechanism, aligning with practical strategies for SNN development and enabling effective decision-making processes.
% Winner-Take-All Mechanism: 
The effectiveness of these proposed structural enhancements is evaluated by integrating high-level models into the BindsNET framework. By leveraging BindsNET, we model the online training of an SNN, integrating the novel structures into the learning process. To ensure the robustness and functionality of our circuits, we employ JoSIM for circuit parameter extraction and functional verification through simulation. 
\end{abstract}


\section{Introduction}
% Intro
\noindent
Growing demand for neural networks has led to innovative solutions that combine fundamental biological principles with hardware implementations. These solutions and the developments in computational neuroscience have a notable influence on the paradigm shift from artificial neural networks (ANNs) to the domain of spiking neural networks (SNNs) due to their distinctive properties of energy efficiency and inference capabilities \cite{towardsSNNjaiswal2019}. SFQ circuits with spike-based behavior show great promise in efficient and fast SNN implementation. 

% Although conventional artificial neural networks (ANNs) utilize continuous-valued functions and supervised (gradient descent-based) learning rules, SNNs exploit sparse neuron spikes with unsupervised learning that can result in lower accuracy and complicated training implementation. Activation functions or neuron models play a pivotal role in the overall learning accuracy and energy efficiency of both ANN and SNN implementations.

% Motivation
Neural data represented with spikes intrinsically resembles the data on superconductor devices \cite{likharev1991a, razmkhahBook}. Furthermore, the shift from the conventional floating point representation to a binary paradigm of 0s and 1s results in notable simplifications and reduced memory requirements. The inherent sparsity in SNNs, where neurons spend most of their time resting, aligns perfectly with the concept of event-driven processing with asynchronous superconductor circuits. This ultimately leads to substantial power savings by eliminating the need for most of the computational operations. Thus, the utilization of superconductor devices on SNN holds great promise for the performance of neuromorphic computing systems \cite{surveySuperconductor}.

% Problem Definition
Superconductor-based SNN designs necessitate the integration of superconductor circuits that accurately replicate the intricate dynamics observed in biological neurons, specifically translating states and actions into neural spikes. Within this paradigm, leveraging the unsupervised learning mechanisms inherent to SNNs, in conjunction with the capabilities of superconductor technology, empowers us to establish a biologically plausible framework for simulating neural networks.
% Moreover, this innovative approach provides an efficient hardware solution for training and executing inference tasks within SNNs. Consequently, the circuits must emulate the functionalities of neurons while remaining adaptable and scalable to meet evolving requirements.
 
Schneider et al. \cite{schneiderNetwork2017} showcased character recognition using an SNN model, explicitly emphasizing the letters 'z,' 'v,' and 'n.' In their study, the authors employed a 3 × 3 input pixel array and implemented a two-layer inference SNN that incorporated Integrate-and-Fire (IF) neurons and Magnetic Josephson Junctions (MJJs).
Bozbey et al. \cite{bozbeyNetwork2020} extended the SNN research by utilizing superconductor Leaky Integrate-and-Fire (LIF) neurons with CMOS-superconductor synapses for the inference SNN. The training process was executed using genetic algorithms applied to the iris dataset. Furthermore, Zhang et al. \cite{zhangNetwork2023} contributed to the field by exploring SNNs featuring IF neurons with Quantum Phase-Slip Junctions (QPSJ). They analyzed superconductor SNN training, using the digit '0' from the MNIST dataset as their experimental basis.
Of particular interest, Segall et al. \cite{stdpSuperconductor} introduced a 1-bit resolution Spike-Timing-Dependent Plasticity (STDP) structure, advancing the prospects of unsupervised learning with superconductor devices.

Collectively, these papers share a common focus on alternative inference neural networks. However, it is crucial to acknowledge a significant limitation—the immaturity of superconductor fabrication technologies for MJJs and QPSJs. Consequently, our contributions primarily center around online training with conventional superconductor elements that can be readily fabricated using available foundry processes.

% Our Contributions
This work focuses on training SNNs utilizing Spike-Timing-Dependent Plasticity (STDP) while providing justifications for integrating superconductor components. Within the scope of our research, we have carefully designed an STDP mechanism tailored for a synaptic finite state machine specifically optimized for Single Flux Quantum (SFQ)--based SNNs. Additionally, we introduce leaky integrate-and-fire (LIF) neurons \textit{with dynamic thresholds} achieved through self-inhibition. This unique feature empowers LIF neurons to adapt dynamically to input patterns, leveraging the temporal diversity among neurons to enhance overall network performance.
% To verify the behavioral characteristics of our circuits, w
We conducted simulations using JoSIM \cite{delportJoSIM} to ensure the functionality and accuracy of our designs.

For network analysis, we leveraged the BindsNET framework's capabilities \cite{bindsnet} and applied them to an architecture representing an asynchronous SNN with two layers \cite{architectureDiehl}. Throughout our analysis, we maintained an evaluation range aligned with the capabilities of superconductor hardware, yielding high levels of accuracy in our observations and assessments.

The key contributions of this paper are as follows. (i) Quantized STDP Mechanism: We introduce a novel STDP mechanism that is quantized and utilizes a leaky non-destructive readout (NDRO); (ii) Dynamic Threshold Behavior: We demonstrate an innovative self-inhibition technique that temporarily modulates LIF neurons' membrane potential, enabling dynamic threshold behavior; (iii) Winner-Take-All Superconductor Structure: We present a novel superconductor structure designed to implement the winner-take-all principle within the context of neural networks; and (iv) Computational Framework: We employ plausible mechanisms within a computational framework to systematically verify and observe the computational behavior of SNNs.

\section{Methodology}
\noindent
The development of spiking neurons and their computational models are dedicated to faithfully mirroring the behavior of biological neurons. These methodologies focus on capturing spike-based activity, which enables precise temporal information encoding. By integrating these neurons with synaptic plasticity mechanisms, neural networks evolve into powerful tools for facilitating unsupervised learning. To achieve this, Spike-Timing-Dependent Plasticity (STDP) plays a key role, enabling networks with adaptive capabilities by modulating the strengths of synaptic connections based on the precise timing of spikes. In the following subsections, we delve into the network architecture and the crucial role of synaptic adaptability as a foundational feature, effectively emulating the intricate dynamics of biological neural networks.

\subsection{Spiking Neural Network Architecture}
\noindent
The network architecture that we follow fundamentally consists of two layers: an input layer and a processing layer as an output layer \cite{architectureDiehl}.
% Figure \ref{fig:architectureOriginal} visually represents the architecture.
For the input, neural encoding techniques are applied to transform input pixels into spikes, such as rate coding, temporal coding, and sparse coding. In particular, we focus on the rate coding scheme that converts a pixel value into a rate of spikes using Poisson distribution \cite{poissonGuo}.
% , thereby adjusting the spike intensity based on the pixel value \cite{poissonGuo}.
In this approach, the source of the input spikes can be any asynchronous input, such as a sensor. The incoming spikes are then propagated to the processing layer after being weighted by synapses.

% \begin{figure}[!t]
% \centering
% \begin{subfigure}{1\linewidth}
%     \centering
%     \includegraphics[width=0.76\linewidth]{networkConfig-01.png}
%     \caption{Spiking neural network architecture with winner-take-all (WTA) principle.}
%     \label{fig:architectureOriginal}
% \end{subfigure}
% \hfill
% \begin{subfigure}{1\linewidth}
%     \centering
%     \includegraphics[width=0.9\linewidth]{RateCoding_Lateral_Mixed-01.png}
%     \caption{\textcolor{black}{The visualization of rate coding on the MNIST input image pixels. In the processing layer, a single excitatory neuron performs lateral inhibition over the other excitatory neurons, preventing them from firing due to the WTA feedback mechanism.}}
%     \label{fig:lateralInhib}
% \end{subfigure}
% \caption{The representation of the original network architecture and its slightly changed version. The labels \textit{In}, \textit{E}, and \textit{I} represent input, excitatory, and inhibitory neuron vertices, respectively. The variable \textit{K} corresponds to the number of input pixels, whereas \textit{N} shows the number of neurons in the processing layer.}
% \label{fig:architecture}
% \end{figure}

\begin{figure}[!t]
\centering
% \begin{subfigure}{1\linewidth}
%     \centering
%     \includegraphics[width=0.76\linewidth]{networkConfig-01.png}
%     \caption{Spiking neural network architecture with winner-take-all (WTA) principle.}
%     \label{fig:architectureOriginal}
% \end{subfigure}
% \hfill
% \begin{subfigure}{1\linewidth}
    \centering
    \includegraphics[width=0.9\linewidth]{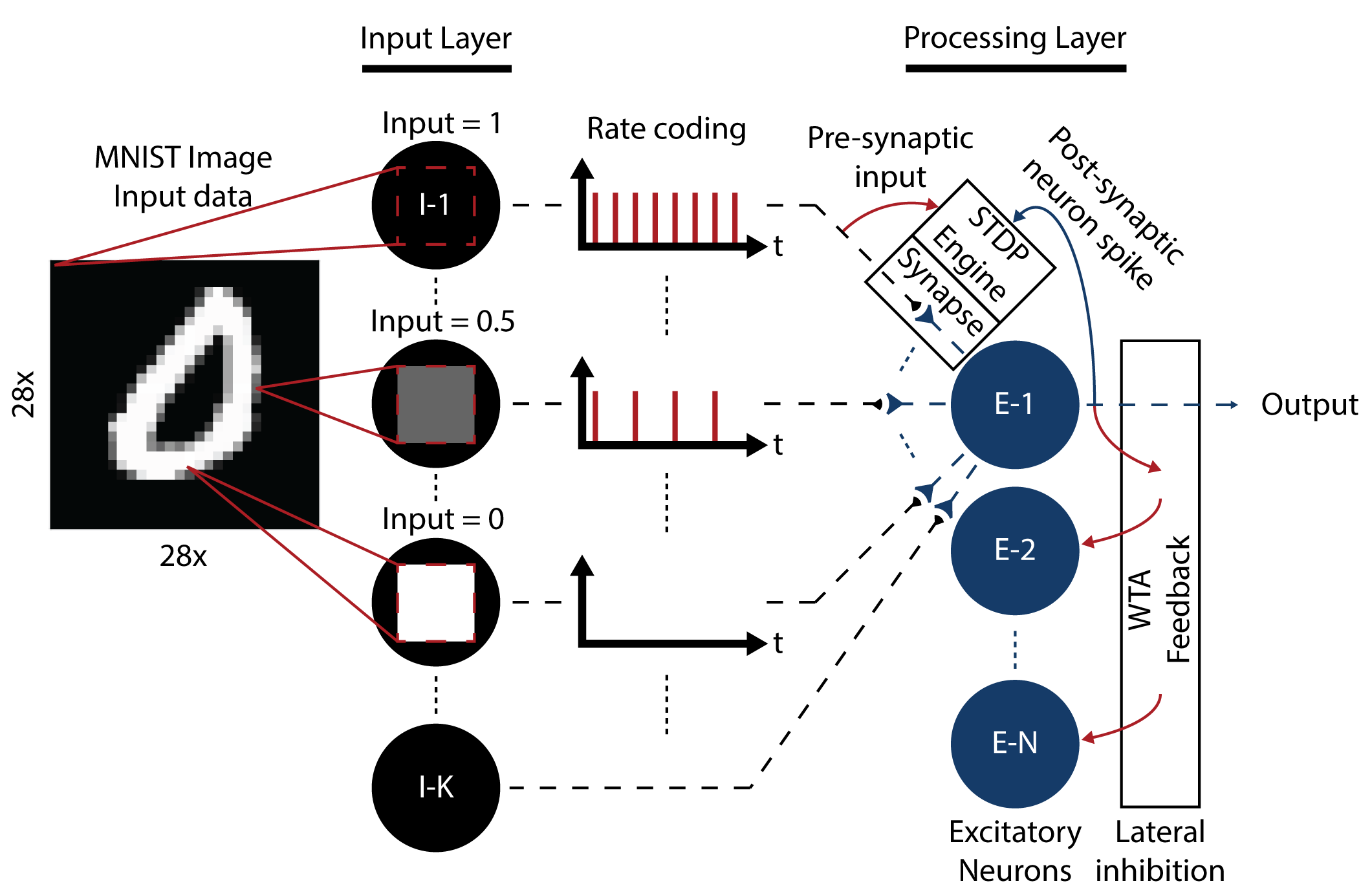}
    \caption{\small\textcolor{black}{The visualization of rate coding on the MNIST input image pixels and our network architecture. In the processing layer, a single excitatory neuron performs lateral inhibition over the other excitatory neurons, preventing them from firing due to the WTA feedback mechanism.} The labels \textit{I} and \textit{E} represent input and excitatory neuron vertices, respectively. The variable \textit{K} corresponds to the number of input pixels, whereas \textit{N} shows the number of neurons in the processing layer.}
    \label{fig:lateralInhib}
% \end{subfigure}
% \caption{The representation of the original network architecture and its slightly changed version. The labels \textit{In}, \textit{E}, and \textit{I} represent input, excitatory, and inhibitory neuron vertices, respectively. The variable \textit{K} corresponds to the number of input pixels, whereas \textit{N} shows the number of neurons in the processing layer.}
\label{fig:architecture}
\end{figure}

The processing layer consists of excitatory and inhibitory neurons. The overall decision-making is performed by excitatory neurons with the help of inhibitory neurons. After being weighted by synapses, the input spikes are initially provided to the excitatory neurons. The synaptic connections from the input layer to excitatory neurons are established in a fully connected fashion. Inhibitory neurons are incorporated into the structure to introduce competitive dynamics among the excitatory neurons.

The connections from excitatory to inhibitory neurons are one-to-one.
% , as given with a dashed blue arrow in Fig. \ref{fig:architectureOriginal}.
In contrast, the links from inhibitory to excitatory neurons are fully connected.
% , represented with red arrows.
Here, the excitatory neuron that provides the initial spikes to the inhibitory neuron is excluded. In this approach, if an excitatory neuron generates output, these spikes trigger the corresponding inhibitory neuron. Once the inhibitory neuron generates an output spike, it will prevent the rest of the excitatory neurons from firing. This paradigm is defined as the winner-take-all (WTA) principle \cite{wtaPrinciple}. This network configuration resembles recurrent neural networks due to lateral inhibition.

In our approach, we utilize a slightly modified version, shown in Fig. \ref{fig:lateralInhib}, of the previously described architecture. For our work, we assigned a single-spike threshold to inhibitory neurons. As a result, these neurons perform just the propagation of spikes with a high fanout back to the excitatory neurons for the operation of inhibition. Therefore, we exclude the inhibitory neurons from the architecture and create a WTA feedback mechanism among the excitatory neurons to establish the same functionality as inhibitory neurons.

\subsection{Spike-timing-dependent plasticity (STDP)}
\noindent
STDP is a phenomenon in which the timing of spikes in neural networks influences both the direction (sign) and magnitude of changes in synaptic strength. It is considered one of the primary learning rules governing synaptic plasticity and is a biologically plausible mechanism for unsupervised learning, as discussed in reference \cite{stdpAsymmetricAbbott}. Conceptually, STDP is often interpreted as a form of Hebbian learning, which posits that synapses are strengthened when neurons fire together.

In STDP, the precise timing of pre-synaptic and post-synaptic neuron spikes within a narrow time window plays a critical role in determining the direction of synaptic changes. When a pre-synaptic neuron spike precedes a post-synaptic neuron spike within this window, it leads to a phenomenon known as long-term potentiation (LTP), which strengthens the synaptic connection. Conversely, if the order is reversed, with the post-synaptic neuron spike preceding the pre-synaptic one, it results in long-term depression (LTD), which weakens the synaptic connection.

In experimental settings, researchers often repeatedly evoke pairs of pre-synaptic and post-synaptic spikes with a fixed time interval, denoted as $\Delta t$. These pairs of spikes are typically repeated at a low frequency, and the resulting changes in synaptic response size are measured. By conducting this experiment for various values of $\Delta t$,  the timing-dependence of plasticity is mapped, creating what is referred to as an \textit{STDP curve}. This curve is a valuable tool for predicting the plasticity outcomes when $\Delta t$ varies, such as in response to arbitrary sequences of pre-synaptic and post-synaptic neuron spikes under less controlled conditions \cite{Shouval2010}. 

\begin{figure}[!t]
\centering
\begin{subfigure}{0.56\linewidth}
    \centering
    \includegraphics[width=\textwidth]{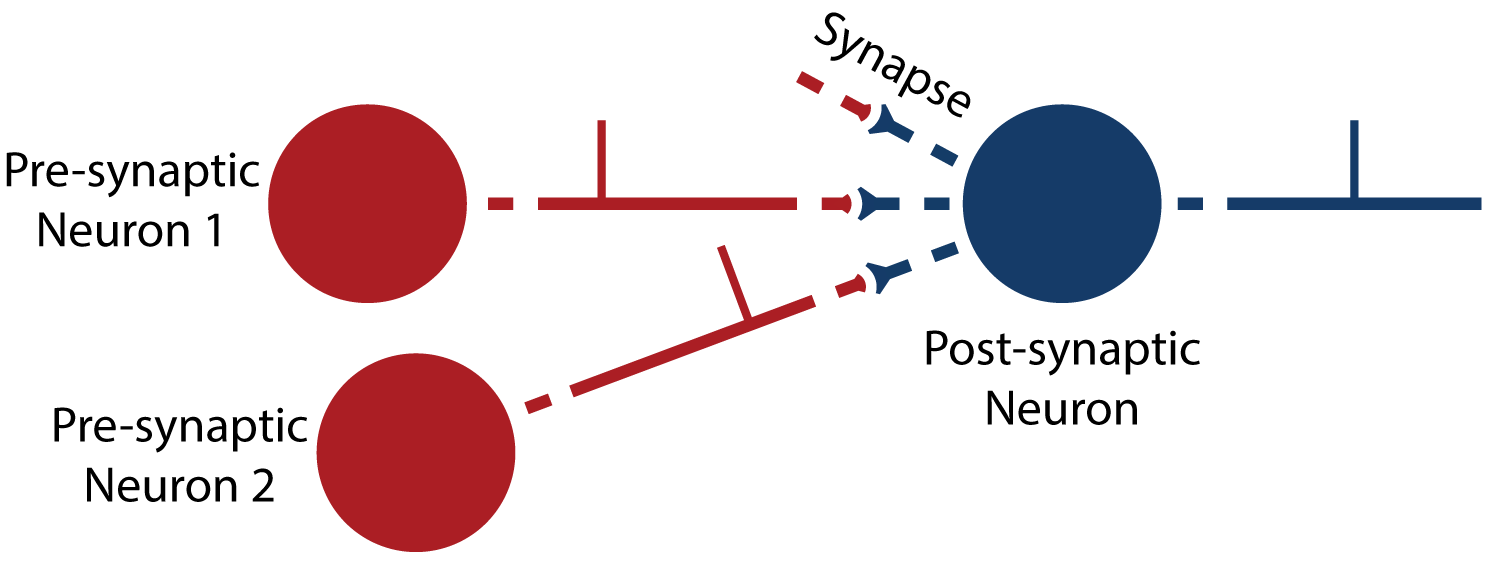}
    % \caption{Example neuron and synapse scheme.}
    \label{fig:neuronSynapse}
\end{subfigure}
\hfill
\begin{subfigure}{0.40\linewidth}
    \centering
    \includegraphics[width=1\textwidth]{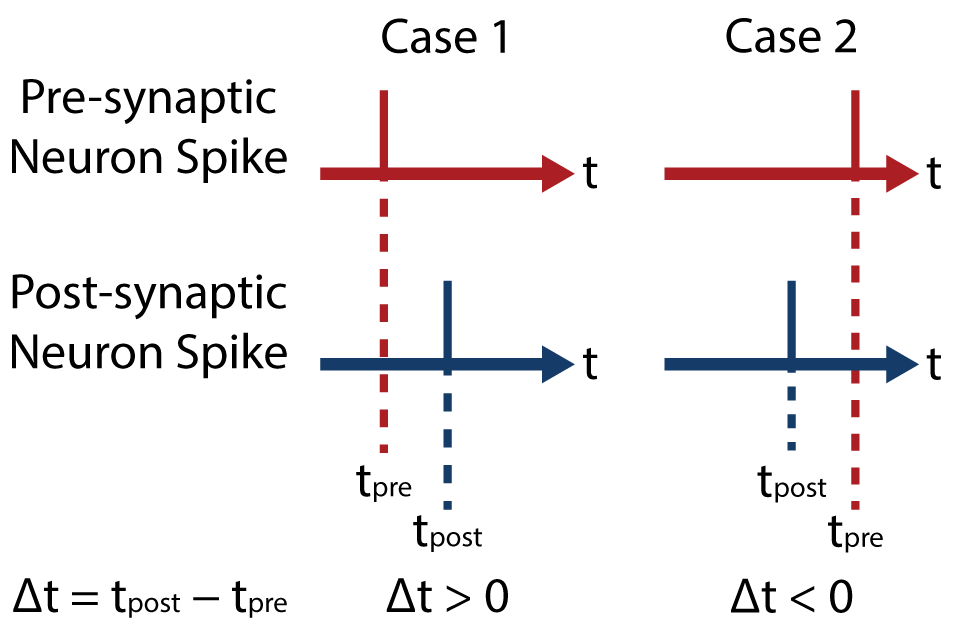}
    % \caption{Pre- and post-synaptic neuron spike cases.}
    \label{fig:stdpTiming}
\end{subfigure}
\hfill
\begin{subfigure}{1\linewidth}
    \centering
    \includegraphics[width=0.66\linewidth]{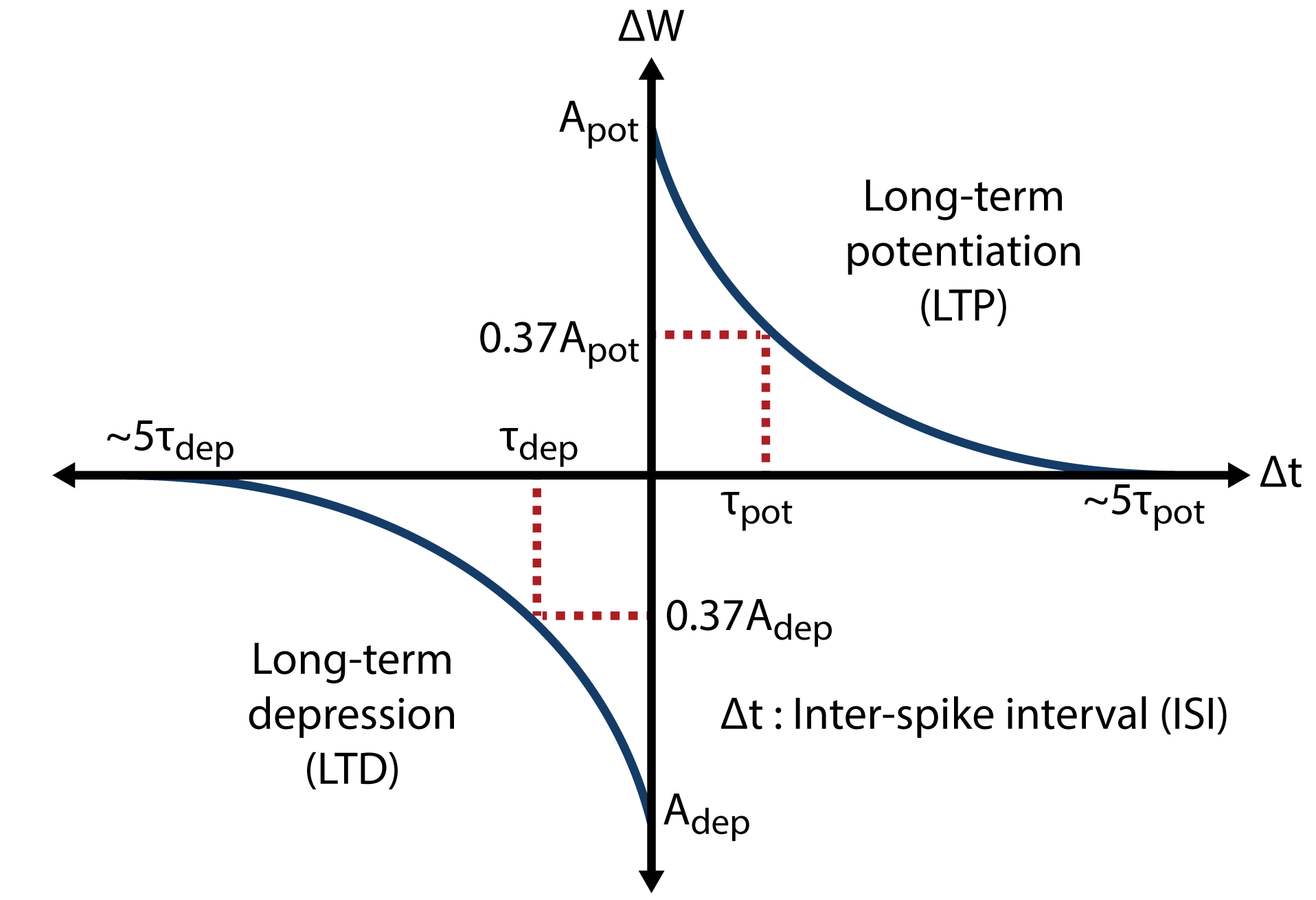}
    % \caption{Learning curves for STDP.}
    \label{fig:stdpCurves}
\end{subfigure}
\caption{\small Example neuron and synapse scheme with pre and post-synaptic neuron spike cases for the learning curves of STDP.}
\label{fig:neuronSynapse_stdp}
\end{figure}

The visual representation of neurons and synapses undergoing different weight update scenarios in the context of STDP is depicted in Fig. \ref{fig:neuronSynapse_stdp}. Let's consider two specific cases. \textbf{Case 1:} In this scenario, the input from pre-synaptic neuron 1 triggers the post-synaptic neuron to generate output spikes. This causal relationship increases the synapse's strength that connects the pre and post-synaptic neurons.
\textbf{Case 2:} In contrast, pre-synaptic neuron 2 does not contribute to the output generation, and its spike arrives later than the output of the post-synaptic neuron. As a result, the strength of the synapse connecting pre-synaptic neuron 2 and the post-synaptic neuron is decreased.

% \textcolor{blue}{What are the definitions of  A+, A-, tau+ and tau-?  The first rule of writing a paper is to define everything.  }

The mathematical expression for the weight changes in these scenarios is provided by Eq. \ref{eq:weightChange}. This equation quantifies how the synaptic weights are updated based on the timing and causal relationship between pre-synaptic and post-synaptic spikes, reflecting the principles of STDP.
\begin{equation} \label{eq:weightChange}
\begin{split}
 \Delta W = \left\{ 
\begin{array}{ c l }
    A_{pot} e^{-\Delta t/\tau_{pot}} & \quad \textrm{if } \Delta t > 0 \\
    A_{dep} e^{+\Delta t/\tau_{dep}} & \quad \textrm{if } \Delta t < 0
\end{array} \right.
\end{split}
\end{equation}

The weight modulation in Eq. \ref{eq:weightChange} expresses the principles of the asymmetric learning rule with two regions: LTP for weight increment and LTD for weight reduction. $\Delta W$ represents the amount of change in synaptic strength. In order to realize this functionality in a circuit, a trace-based method can be implemented \cite{stdpTraceMorrison2008}. The exponential curve on the LTP region corresponds to the post-synaptic neuron spike-trace, whereas the LTD region curve represents the pre-synaptic neuron spike-trace. These traces capture the spiking activity of the pre-synaptic and post-synaptic neurons. The two variables $A_{pot}$ and $A_{dep}$ denote the maximum increment and decrement of synaptic strength on the \textit{STDP curve}, respectively. The direction of $\Delta W$ depends on the sign value of $\Delta t$ as determined by the arrival order of pre and post-synaptic neuron spikes corresponding to $\Delta t = t_{post} - t_{pre}$. Due to the resource constraints on the superconductor hardware, we employed quantized STDP update levels on $\Delta W$ and synaptic weights that are discussed in the following section.

% During the weight update process, \textcolor{blue}{What do you mean by "pre and post-traces on the learning curve"? Where did you define them previously? Where are they in Figure 4, which is your learning curve?} The pre and post-traces on the learning curve can be \textcolor{blue}{what is being discretized? a trace? what is a trace? A path cannot be a scalar value. What does it mean to discretize a hint, then?} discretized to lower bit-resolution, as shown with a horizontal line below the exponential decay.

% \textcolor{blue}{How is this engine instantiated in the context of pre and post-synaptic neurons? You must relate Figure 5 to what is shown in figure 3.  Is this engine a replacement for the synaptic connection?}

% \textcolor{blue}{How many unsupervised learning rules are there? you just mentioned one i.e., STDP? Are there more rules you are using?} 

\section{Proposed SFQ-based Online Training}
\noindent
This section introduces essential superconductor-based mechanisms for the implementation of SNNs. Firstly, we present an STDP engine that effectively enforces the asymmetric rate STDP learning rule, underlining the motivation for synaptic implementations with a degree of biological plausibility \cite{stdpAsymmetricAbbott}. Our design employs a \textit{modified NDRO circuit} to monitor pre- and post-synaptic neuron spikes precisely.
Furthermore, our framework seamlessly integrates dynamic threshold behavior within LIF neurons, achieved through self-inhibition based on neuron output spikes \cite{selfInhibitionLiu2001}.
Lastly, we propose a novel implementation of feedback interactions between neurons, facilitating our framework's realization of WTA characteristics. These combined mechanisms advance the field of SNNs and provide valuable tools for neuromorphic computing applications.

% \begin{figure}[!t]
% \centering
%     \includegraphics[width=0.9\linewidth]{modifiedArch.png}
%     \caption{Representation of modified architecture with the proposed designs. \textcolor{red}{Altay: I will redraw or add it to Fig 1b}}
%     \label{fig:modifiedArch}
% \end{figure}

\subsection{SFQ-based STDP Engine}
\begin{figure}[!t]
\centering
    \includegraphics[width=0.76\linewidth]{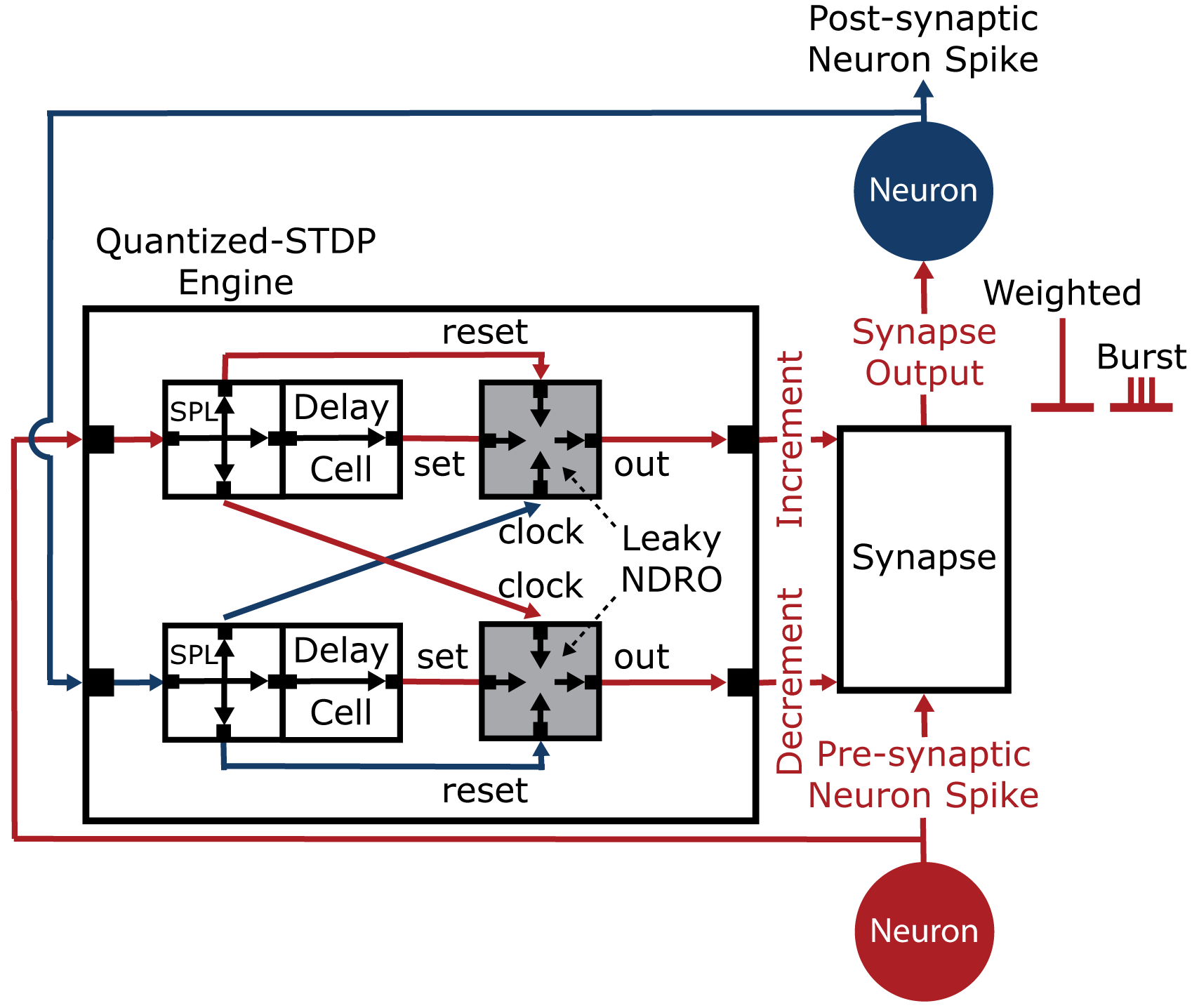}
    \caption{\small Cell view of the quantized STDP design following the scheme given in Fig. \ref{fig:neuronSynapse_stdp}. Trace-based spiking activity of pre and post-synaptic neurons is recorded in each leaky NDRO.}
    \label{fig:controllerSTDP}
\end{figure}

\noindent
The proposed design of the STDP mechanism is tailored to perform both increment and decrement functions, thereby adjusting a finite state machine associated with the synapse structure. To enable STDP using superconductor-based components, we discretized the learning curve. This design, featuring a 1-bit resolution, incorporates two splitters (SPLs) with a fanout of 3, in addition to two leaky Non-Destructive Readout (LNDRO) circuits. The SPLs play a crucial role in internally increasing the fanout of inputs derived from pre- and post-synaptic neuron spikes, which are then assigned to the LNDRO pins. Furthermore, the STDP engine includes two output pins, labeled as \textit{increment} and \textit{decrement}, as illustrated in Fig. \ref{fig:controllerSTDP}.

\begin{figure}[!t]
\centering
\begin{subfigure}{1\linewidth}
    \centering
    \includegraphics[width=0.98\linewidth]{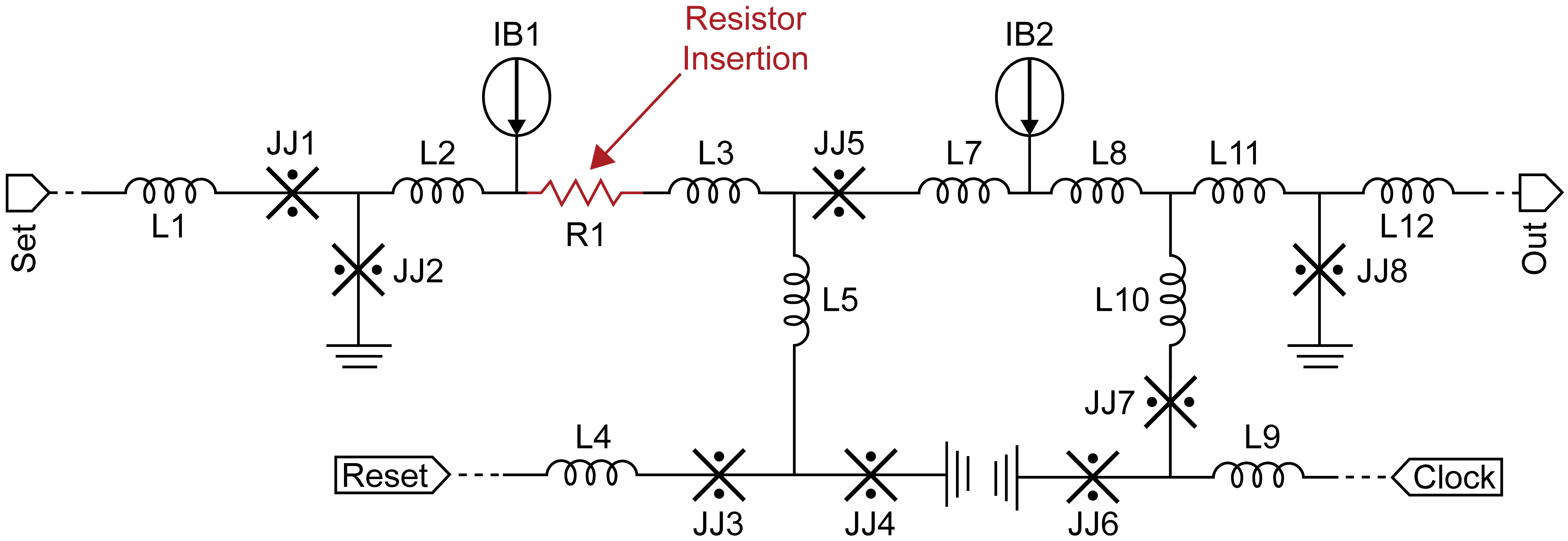}
    \caption{\footnotesize LNDRO schematic.}
    \label{fig:leakyNDROsch}
\end{subfigure}
\hfill
\begin{subfigure}{0.53\linewidth}
    \centering
    \includegraphics[width=1\linewidth]{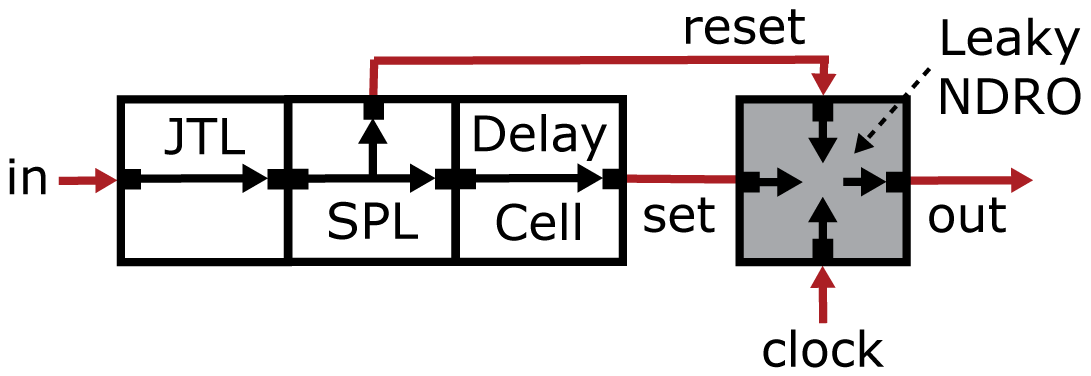}
    \caption{\footnotesize LNDRO testbench.}
    \label{fig:leakyNDROtb}
\end{subfigure}
\hfill
\begin{subfigure}{0.41\linewidth}
    \centering
    \includegraphics[width=1\linewidth]{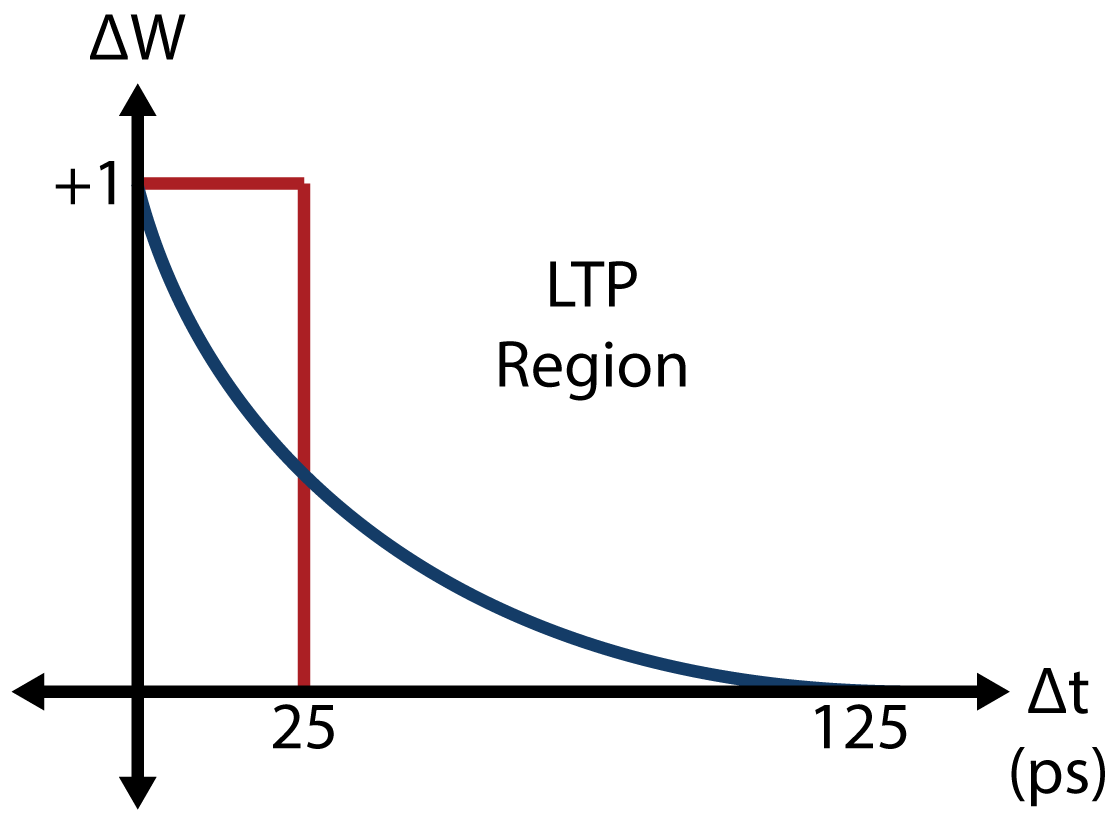}
    \caption{\footnotesize Representation of STDP learning curve on LTP region.}
    \label{fig:stdpQuant}
\end{subfigure}
\hfill
\begin{subfigure}{1\linewidth}
    \centering
    \includegraphics[width=0.74\linewidth]{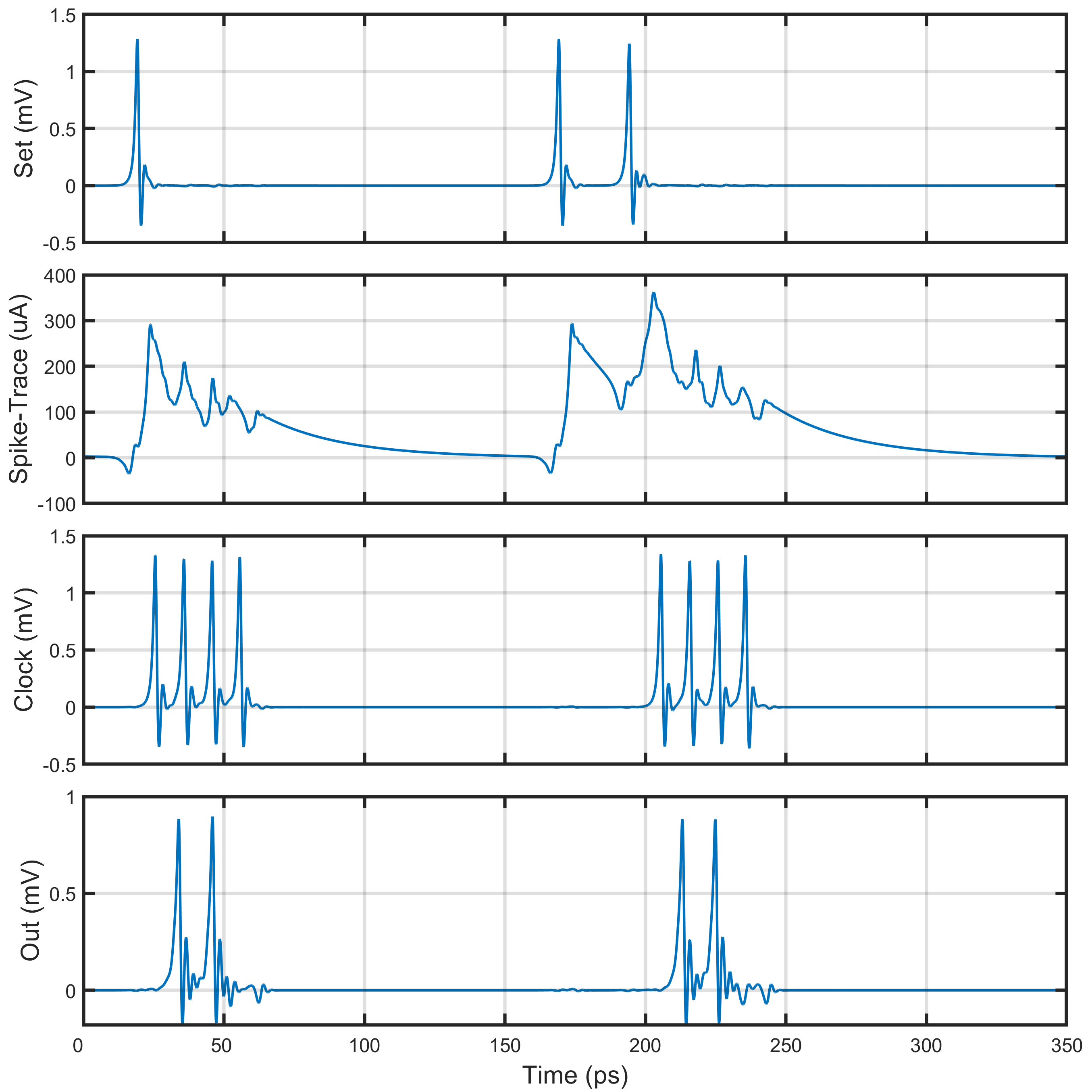}
    \caption{\footnotesize LNDRO JoSIM result. Since a spike on \textit{reset} arrives prior to a spike on \textit{set} to re-initiate the current for decay, the \textit{Reset} waveform is omitted for simplicity. In this simulation, the decay rate $\tau$ for the spike-trace of LTP is set to $\approx$25 ps, whereas the inter-delay of spikes on \textit{clock} is 10 ps. The current decay is observed on the inserted resistor \textit{R1} (0.26 $\Omega$). In the quantized STDP model (the red line in part c), an output spike is generated only when the arrival of a spike from the \textit{clock} aligns within a dedicated time window.}
    \label{fig:leakyNDROsim}
\end{subfigure}
\caption{\small LNDRO functional verification and its representation for the quantized STDP on the learning curve.}
\label{fig:leakyNDRO}
\end{figure}

To generate a decrement output spike, one must establish the spike-trace relationship between pre- and post-synaptic neurons. In this operation, incorporating Splitters (SPLs) and a single LNDRO cell proves sufficient to generate decrement behavior with a 1-bit resolution.
A similar setup can be set up to produce spikes on the increment output pin for generating an increment behavior. If a higher bit resolution is desired, the overall design necessitates the incorporation of SPLs with increased fanout, additional LNDROs with varying decay rates, and the inclusion of two merger components for LNDRO outputs. In this design, an NDRO with multi-flux storing characteristics is suitable to create the synapse behavior since the amount of stored flux corresponds to the state of a synapse \cite{mndroZeynep2023}. 
The introduction of leaky behavior in the standard NDRO cell is achieved by inserting a resistor into the SFQ storage loop, as illustrated in Fig. \ref{fig:leakyNDROsch}.

% \textcolor{blue}{MP -- What does it mean to apply a neuron pulse first to Reset pin and then to set pin. Is the same pulse being applied? Is there a delay introduced between the application of these pulses? Is there a splitter? where is the associated circuitry? The writing is poor and even on a second and third attempt you do not fix it.}
% provided to the \textit{Post} pin (Reset) and then to the \textit{Delayed Post} (Set) pin. 
The LNDRO is simulated using JoSIM, and the waveforms are given in Fig. \ref{fig:leakyNDROsim}. When a spike arrives at the input \textit{in}, the spike is split for the \textit{reset} and \textit{set} pins. To recreate the correct spike-trace functionality, the spike from \textit{reset} initially erases the current in the leaky storing loop (JJ2-L2-R1-L3-L5-JJ4). By applying the spike arriving late due to the \textit{Delay cell}, the spike-trace is updated, and the current gradually decays with a constant time of $\tau$ due to the inserted resistor.
% \textcolor{blue}{MP -- which loop are you talking about? The one that contains R1? 
% Specify the loop in Fig. 6}
% The loop current introduced by a pulse from \textit{Delayed Post} gradually decays with a constant time of $\tau$ due to the inserted resistor. 
Concurrently, the state of the LNDRO can be read by a spike from the \textit{clock} pin. Due to the quantization of the STDP curve shown in Fig. \ref{fig:stdpQuant}, a spike on \textit{out} can only be observed until a quantization point, set as 25 ps in the simulation.
% \textcolor{blue}{MP -- where do I find the phrase "Post-Trace" in these figures? The paper is not written well.} \textcolor{red}{Altay: Professor it was in the simulation figure. I renamed everything now. Will change the text accordingly.}

\subsection{LIF Superconductor Neurons with Dynamic Threshold}
\begin{figure}[!t]
\centering
    \includegraphics[width=1\linewidth]{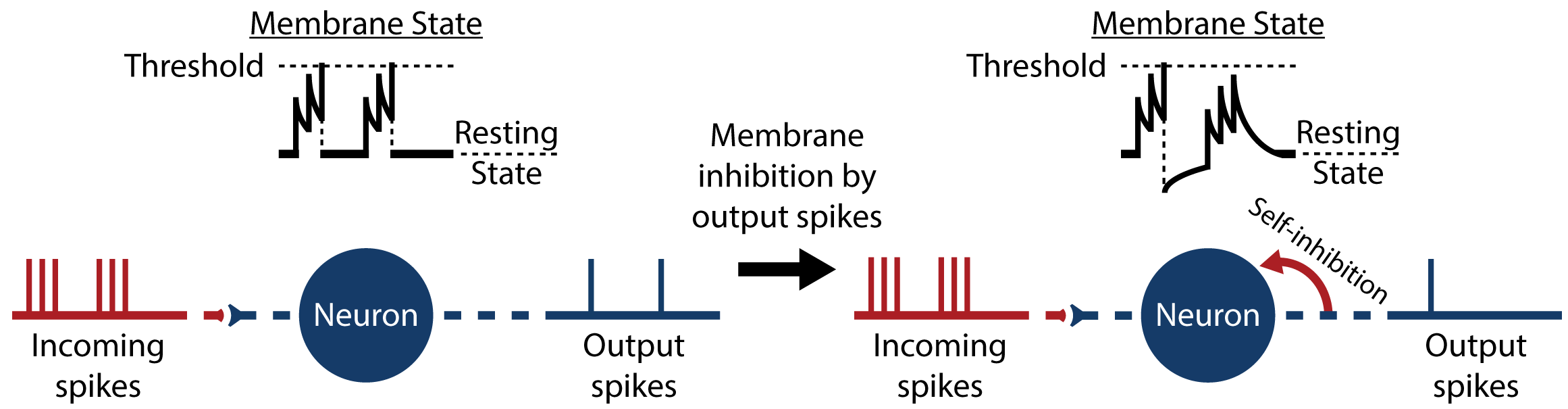}
    \caption{\small LIF neuron with dynamic threshold using self-inhibition. The membrane state of a superconductor LIF neuron corresponds to the amount of current stored in its leaky loop. The neuron resting state is the baseline condition where no spike is received or generated.}
    \label{fig:neuronSelfInhib}
\end{figure}

\noindent
The threshold value of a neuron determines the firing rate and shapes the computational behavior of a neural network. In this context, relying on a neuron with a fixed threshold can impose challenges in neural processing, such as high sensitivity to input fluctuations and excessive spike firing, resulting in high dynamic power consumption. The adaptability of a neuron with a dynamic threshold contributes to the network stability and contextual responsiveness \cite{adaptiveThreshold}. For instance, the digits in the MNIST handwritten dataset may share the same pixels. In this case, the multiple neurons in the output layer may have high-valued synaptic weights on these pixels and generate an output spike due to the neurons having the same threshold for the classification, hindering the overall performance. Hence, the implementation of an adaptive threshold behavior becomes indispensable to address this issue effectively.
Adjusting the threshold of a superconductor-based neuron is typically achieved by dynamically changing the bias current and critical current of the JJs. Such modifications introduce additional hardware design complexity. Therefore, we utilize the generated output spike as a feedback input, creating self-inhibition behavior as shown in Fig. \ref{fig:neuronSelfInhib}.

The membrane state of a neuron plays a pivotal role in determining whether an output spike is generated at a specific moment in time. When the membrane state surpasses a certain threshold, a single spike is triggered at the neuron's output. The resting state of a neuron corresponds to its default state, characterized by the absence of spike activity. Within a neural network, certain neurons may exhibit a high firing rate, exerting a disproportionate influence on decision-making and potentially disrupting the network's balance. Consequently, there is a need for methods to mitigate such behavior and maintain network equilibrium.

\begin{figure}[!t]
\centering
\begin{subfigure}{1\linewidth}
    \centering
    \includegraphics[width=1\linewidth]{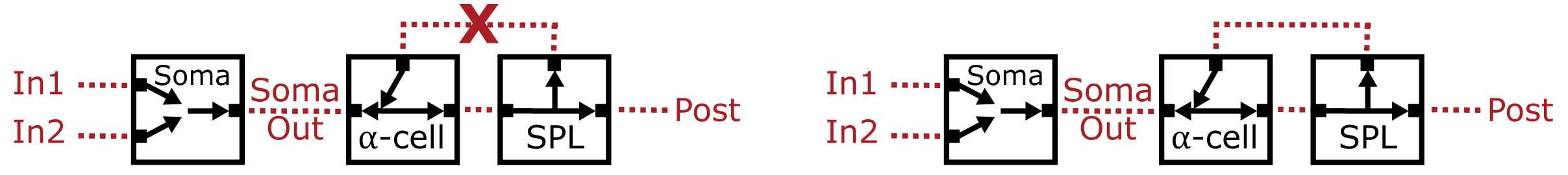}
    \caption{\footnotesize Cell view for LIF neuron behavior with $\upalpha$-Soma using the same cell parameters (left without self-inhibition, right with self-inhibition).}
    \label{fig:alphaSelfInhib}
\end{subfigure}
\hfill
% \begin{subfigure}{1\linewidth}
%     \centering
%     \includegraphics[width=0.8\textwidth]{alphaSomaLateralNo.png}
%     \caption{Simulation without self-inhibition.}
%     \label{fig:alphaSelfInhibSim}
% \end{subfigure}
% \hfill
\begin{subfigure}{1\linewidth}
    \centering
    \includegraphics[width=0.74\textwidth]{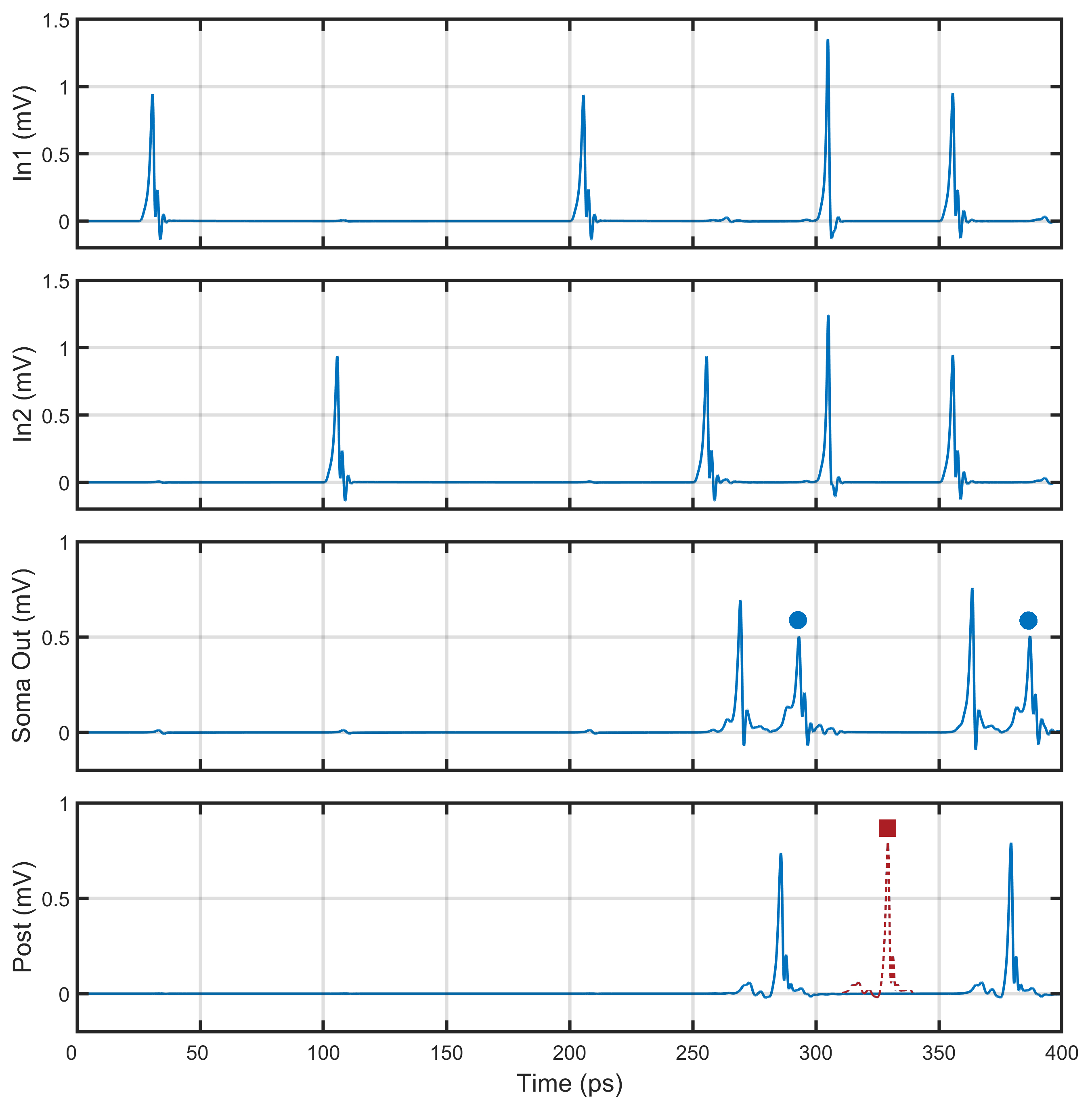}
    \caption{\footnotesize Simulation with digital self-inhibition. The self-inhibition spikes are marked with a circle. In the case of no self-inhibition, the spike with a square mark would also be expected.}
    \label{fig:alphaSelfInhibSim}
\end{subfigure}
\caption{\small Design of LIF neuron with self-inhibition and JoSIM results for $\upalpha$-Soma. All parameter values of the cells are kept the same, and the self-inhibition is prevented by disconnecting the node between the input of $\upalpha$-cell and output of SPL.}
\label{fig:alphaSelf}
\end{figure}

The proposed self-inhibition technique serves as a mechanism for temporarily preventing a neuron from firing multiple spikes in quick succession. It achieves this by reducing the membrane state below the resting state, as illustrated in Fig. \ref{fig:neuronSelfInhib}. In the context of superconductor circuits, the membrane state corresponds to the amplitude of stored current, and the threshold is determined by the amount of current required to trigger the decision-making Josephson junction (JJ). By employing the self-inhibition mechanism, a neuron can momentarily dip below its resting state, thus demanding more input current to reach the threshold level. An example design utilizing $\upalpha$-Soma \cite{alphaSoma} is depicted in Fig. \ref{fig:alphaSelfInhib}.
A similar effect can be achieved for neurons receiving inputs from inductances with mutual couplings \cite{kenSegallFaninFanout2020} by directing the neuron's output spike back into the same neuron's input through an inductance with negative coupling. This generates a current in the opposite direction, counteracting the threshold mechanism.
% Note that the decay time of the entire structure is shared since all inputs are accumulated within the same current storing loop. To create distinct decay times, an additional input path with a different decay rate has to be established for the same neuron.

The simplified test case utilizes a soma circuit that acts as an asynchronous threshold gate with a threshold of two spikes within $\approx$50 ps i.e., an output spike is generated whenever two input spikes arrive within the designated time frame. Within this configuration employing $\upalpha$-Soma, the output spike is provided to the following SPL cell. One of the SPL outputs is connected to the input of $\upalpha$-cell, enabling the spike to propagate back into the soma cell. This operation negatively impacts the loop current due to the opposite direction of the spike.
% In addition to the case without self-inhibition, t
The example demonstration of digital inhibition with $\upalpha$-Soma preventing the generation of continuous output spikes is shown in Fig. \ref{fig:alphaSelf}.

\subsection{WTA Mechanism with Superconductor Devices}
\noindent
One of the fundamental computational models in spiking neural networks is a winner-take-all (WTA) principle, establishing a form of competition for activation among the neurons within the same layer \cite{wtaPrinciple}. The neuron with a higher activation affects the activity of interconnected neurons by inhibition. As a result, the winner neuron becomes the sole source of output spikes. This selective mechanism enables noise filtering and input focus on the critical data.

We present a way that enables interaction among the excitatory neurons to prevent each other from firing, fundamentally implementing the WTA principle. The firing information is obtained from an inductance at the output \textit{Lload}, next to the JJ, that determines the threshold operation. Note that the choice of inductance for coupling can be placed between the threshold junction \textit{JJsoma1} and the ground node; however, this option requires a balance adjustment between the decay rate of the input side and lateral inhibition since such inductance is a part of the leaky storage loop of the soma circuit \cite{alphaSoma}. 

\begin{figure}[!t]
\centering
\begin{subfigure}{1\linewidth}
    \centering
    \includegraphics[width=0.66\linewidth]{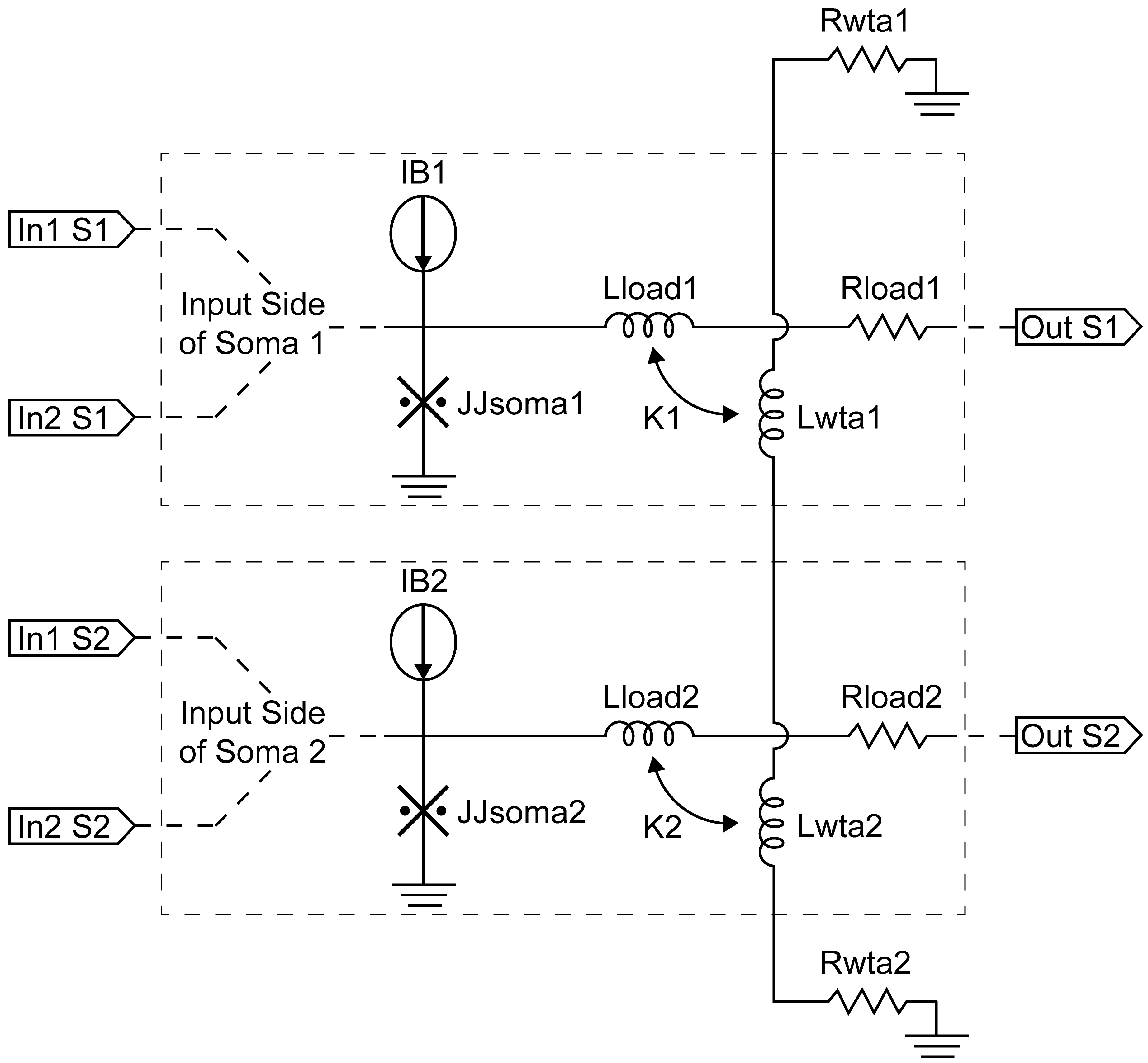}
    \caption{\footnotesize WTA mechanism with superconductor components. S1 and S2 represent the labels for Soma 1 and 2, respectively.}
    % The first firing soma suppresses the other soma with the help of a feedback loop.
    \label{fig:WTAcircuit}
\end{subfigure}
\hfill
% \begin{subfigure}{1\linewidth}
%     \centering
%     \includegraphics[width=0.8\textwidth]{WTAnoCoupling.PNG}
%     \caption{Simulation without WTA mechanism.}
%     \label{fig:wtaNoCouple}
% \end{subfigure}
% \hfill
\begin{subfigure}{1\linewidth}
    \centering
    \includegraphics[width=0.74\textwidth]{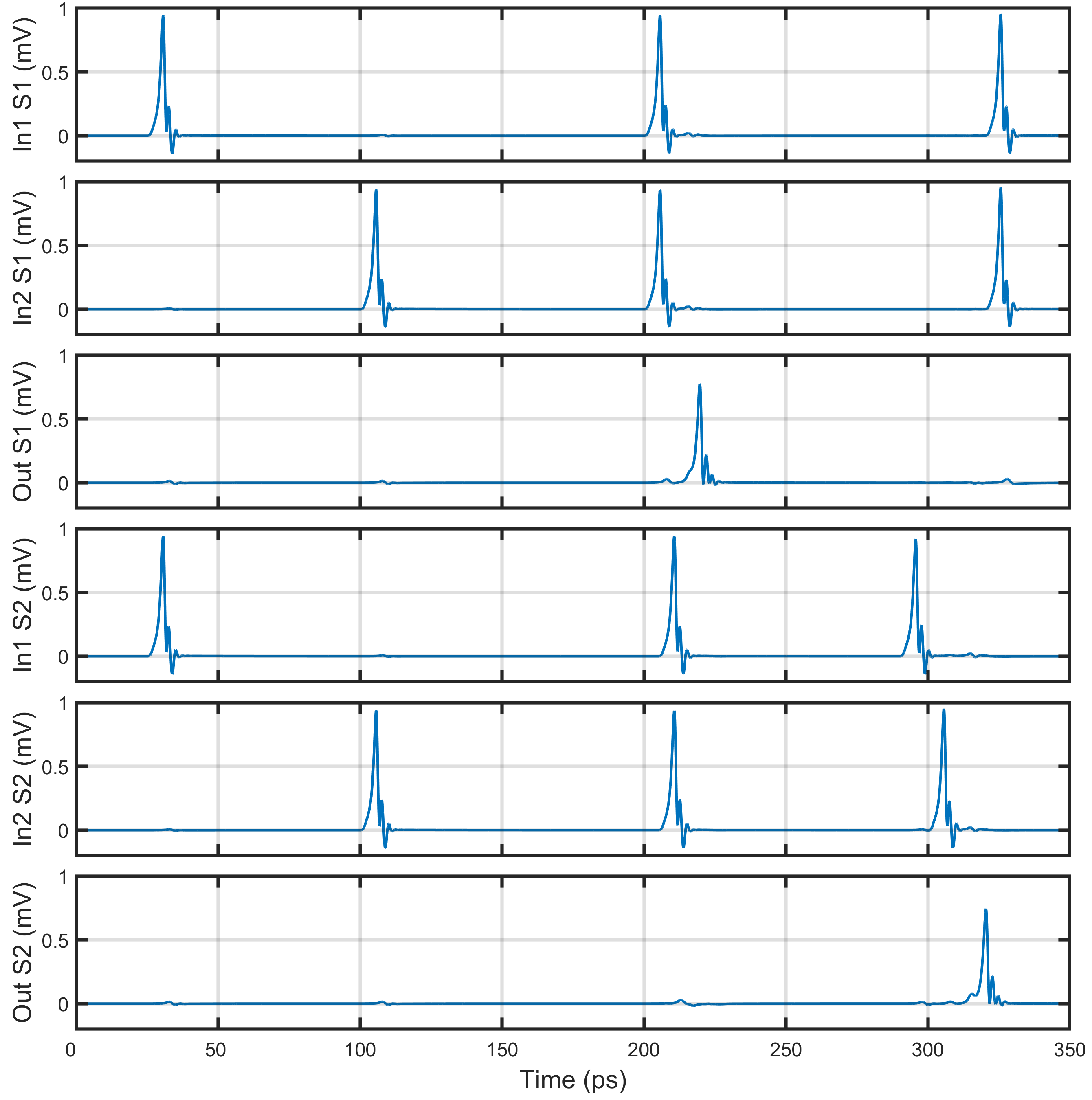}
    \caption{\footnotesize Simulation of WTA mechanism. (Lload = 2 pH, Rload = 0.6 $\Omega$, Lwta = 3 pH, Rwta = 0.3 $\Omega$, K = 0.7)}
    \label{fig:wtaCouple}
\end{subfigure}
\caption{\small WTA mechanism and JoSIM results for soma circuits.}
\label{fig:wtaSIM}
\end{figure}

The interaction among neurons is established by inductive coupling \textit{K} between an output inductance (Lload) and inductance in a feedback loop (Rwta1-Lwta1-Lwta2-Rwta2) shown in Fig. \ref{fig:WTAcircuit}. The feedback loop consists of Lwta inductances for each neuron and resistors (Rwta) on each end. Therefore, any activation on a threshold junction will result in a change of current within this loop. Due to the coupling, the current on the feedback loop will affect the other neurons via coupled inductances. Each input spike creates a current determined by a resistor within the leaky storage loop of superconductor LIF neurons. Unlike this input, the inhibition current from a neuron to other neurons mainly depends on the value of the coupling \textit{K}. Therefore, high inductive coupling results in a higher lateral suppression by a neuron. The JoSIM results of the proposed method with an example of soma circuits are shown in Fig. \ref{fig:wtaSIM}.

In our testbench, two soma circuits are designed to have a threshold of 2 spikes and $\approx$ 50 ps decay time. Initially, these circuits receive spikes with 75 ps time difference. Due to the current decay within the spike-storing loop on the input side of the soma, no output is generated. Next, two spikes with a short time interval are applied to the soma circuits. The first soma receives its inputs before the inputs of the second soma. As a result, the first soma fires earlier than the second one. The late-firing soma is expected to generate an output when no interaction exists among the somas. However, if the WTA mechanism is employed, the first soma inhibits the second soma. Note that there will be no lateral inhibition if the spike frames between the soma circuits do not overlap.

\section{Results}
\noindent
In our work, we utilized the BindsNET framework for high-level network modeling. We primarily focused on the intrinsic network properties and performance. We targeted the digits 0 and 1 of the MNIST dataset to evaluate the on-chip training network. The image count was 633 for the training and 105 for the testing images of 0s and 1s, corresponding to 5\% of the overall digit images of 0s and 1s within the dataset. Images were randomly selected, and each image was down-scaled from 28 by 28 pixels to 14 by 14 pixels. The training epoch count was assigned as 1, and each image was shown for 100 ps to the network.

The neuron resting state was 0, and their neuron threshold was set to eight spikes. While threshold decay to the resting state was active, the dynamic threshold increment from the post-synaptic neuron spike was set to 32 inhibitory spikes. In our network structure, this value prevented the firing neurons from entering burst mode and gave other neurons a chance to fire. The membrane decay time of neurons was set to 25 ps.

For the synaptic characteristics, the increment and decrement in weight adjustments from the STDP mechanism were assigned as two spikes and one spike, respectively. Therefore, the spike generation on the increment side of the proposed STDP engine required a modification to have twice as much impact as the decrement side. For the spike activity of pre- and post-synaptic neurons, we assigned 10 ps to the spike-trace decay rate $\tau$, keeping it within a reasonable time frame. Two network architectures were considered, including architectures with 4 and 9 excitatory neurons with a weight resolution of 4 bits, corresponding to 16 different synaptic stages.

Unlike the conventional implementations, we did not use any weight normalization technique in the on-chip training setting. Accessing all weights to perform such an operation is not hardware-friendly, even though weight normalization gives all neurons a chance to fire and offers a better weight convergence. The dynamic threshold adaptation with decaying fashion temporarily compensates for the weight normalization. Some neurons can rapidly go into a burst mode without this feature and become dominant. However, when the decay time is relatively short compared to the duration of multiple input images, the neuron tends to forget the learned pattern, leading to an increase in the majority of its weights. On the other hand, we kept a weight clipping operation to limit the weight between 0 and 1 (scaled from the range of 0 - 15). For the overall configuration, the estimated JJ count would be $\approx$ 31k with four neurons and $\approx$ 85k with nine neurons. 

% Input neuron: 14x14=156, 156x2=312 JJs
% Excitatory neuron: 14x14=156, (156+1)*4=628 JJs
% STDP: 14x14x4=624, 624x42=26208 JJs
% (8 LNDRO + 2 Delay + 4 SPL) x 2 + (8 from SPL+Merge for inc pin) + 8 NDRO= 42 JJs
% Interconnects: PTL RX + TX = 14x14x4x(3+2)= 3920 JJs

% Input neuron: 14x14=156, 156x2=312 JJs
% Excitatory neuron: 14x14=156, (156+1)*9=1413 JJs
% STDP: 14x14x9=1764, 1764x42=74088 JJs
% (8 LNDRO + 2 Delay + 4 SPL) x 2 + (8 from SPL+Merge for inc pin) + 8 NDRO= 42 JJs
% Interconnects: PTL RX + TX = 14x14x9x(3+2)= 8820 JJs

\begin{figure}[!h]
\centering
    \includegraphics[width=0.9\linewidth]{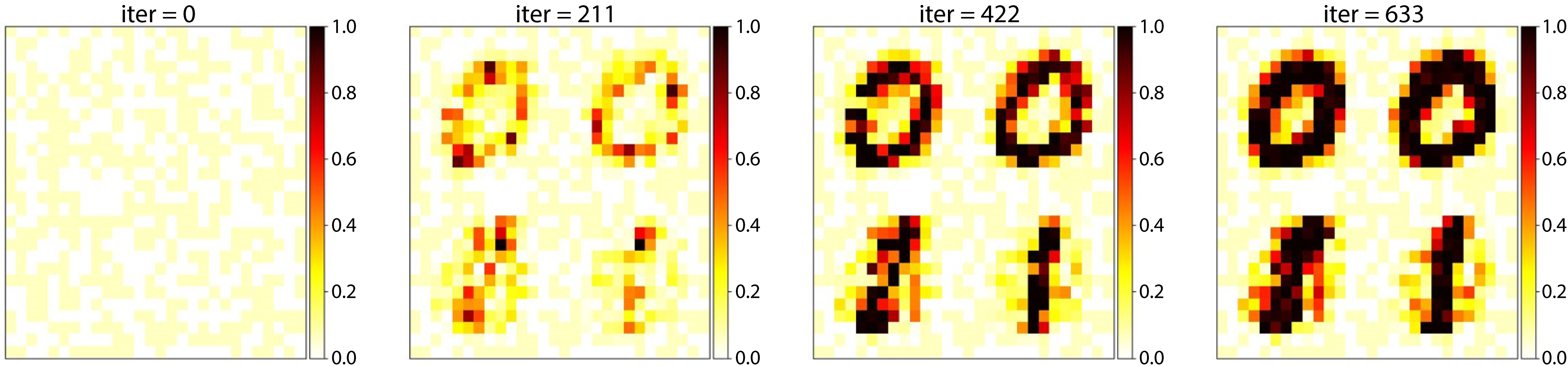}
    \caption{Network training result using four excitatory neurons.}
     % for the classification of 0s and 1s.
    \label{fig:network4neuron1}
\end{figure}

In the case of training first architecture (4 excitatory neurons), we acquired an accuracy of 90.32\% for training and 81.9\% for testing. During the convergence, we observed that the neurons representing digit 1 still have traces of digit 0 and vice versa due to the bit resolution and no weight normalization. We also observed cases where some neurons go into a burst mode due to the random weight initialization. Therefore, a mechanism to thoroughly address this issue in lower-bit resolutions must be investigated. The 2D weight values are illustrated in Fig. \ref{fig:network4neuron1} for each neuron of the considered architecture. The displayed weight values, arranged from left to right, represent snapshots taken at every one-third interval of the training process, spanning from iteration 0 to 633.

\begin{figure}[!h]
\centering
    \includegraphics[width=0.9\linewidth]{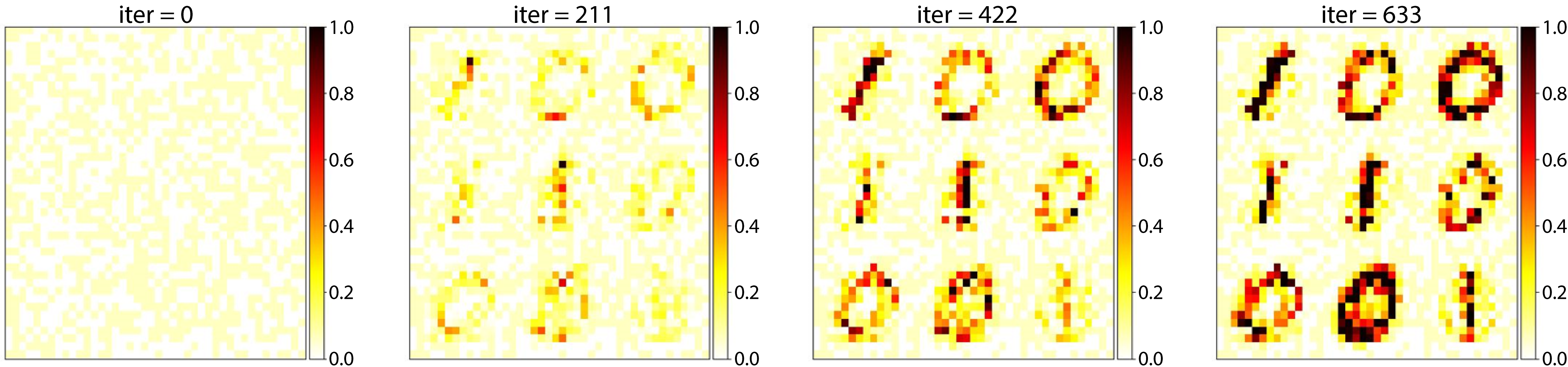}
    \caption{Network training result using nine excitatory neurons.}
     % for the classification of 0s and 1s.
    \label{fig:network4neuron2}
\end{figure}

The training settings are kept the same for the second network architecture (with nine neurons) to observe the impact of neuron count on the performance. The results showed an accuracy of 96.77\% for training and 97.1\% for testing, indicating a performance improvement. Therefore, increasing the number of neurons in the network positively influences the overall accuracy with a trade-off of hardware resources, supporting the motivation for large-scale implementations. The 2D weight values of this architecture during the training are illustrated in Fig. \ref{fig:network4neuron2}

\section{Conclusion}
\noindent
This paper explored the capabilities of an on-chip training mechanism on superconductor spiking neural networks. We designed a leaky NDRO circuit and simulated its behavior with JoSIM. The leaky NDROs record spike traces to achieve a quantized STDP mechanism. Furthermore, we demonstrate a self-inhibition method for superconductor-based structures to establish the dynamic threshold behavior in LIF neurons. We also implement a superconductor winner-take-all mechanism to support the correct network behavior. The on-chip training capabilities are shown with a computational BindsNET framework, and we achieved $\approx$97\% accuracy with 9 neurons for the classification of digits 0 and 1. These findings collectively highlight the promise of on-chip training in superconductor-based spiking neural networks.

%\noindent
%Acknowledgment: This work has been funded by the National Science Foundation (under the project Expedition Discover (Design and Integration of Superconducting Computation for Ventures beyond Exascale Realization) grant number 2124453.

% \bibliographystyle{IEEEtran}
% \bibliography{references}

\end{document}